# What Is A Quantum Really Like?

Ghenadie N. Mardari

*Rutgers University, 89 George St., New Brunswick, NJ 08901*

**Abstract.** The hypothesis of quantum self-interference is not directly observable, but has at least three necessary implications. First, a quantum entity must have no less than two open paths. Second, the size of the interval between any two consecutive quanta must be irrelevant. Third, which-path information must not be available to any observer. All of these predictions have been tested and found to be false. A similar demonstration is provided for the hypothesis of quantum erasure. In contrast, if quanta are treated as real particles, acting as sources of real waves, then all types of interference can be explained with a single causal mechanism, without logical or experimental inconsistencies.

In the early days of quantum mechanics, there were no easy ways to test its interpretations. More often than not, scientists had to rely on thought experiments. In our days, on the other hand, it is hard to come up with a quantum property that is not already testable. Granted, some features of quantum systems are undetectable by definition. However, unobservable causes are only relevant if they produce observable outcomes. Therefore, any hypothesis about quantum properties can be verified by checking its necessary implications against actual experimental data. For example, the concept of self-interference leads to three major predictions. Firstly, a particle must cover more than one path if it is to interfere with itself. This might happen through non-local propagation, or local splitting-recombination. Either way, it must be essential to have two open paths (or more) for every single photon, whenever self-interference is at work. Secondly, the presence or absence of other quanta must be of no consequence. Large numbers of coherent photons are still necessary for the detection of fringes, but the size of the interval between any two consecutive quanta must be irrelevant. They could be light-years away from each other, as far as self-interference is concerned. Thirdly, it is also known that interference vanishes whenever direct path verification is attempted. This is frequently cited as proof that path distinguishability destroys self-interference. If so, any indication about actual photonic trajectories, even if obtained by non-invasive means, should be incompatible with the observation of fringes in the same experiment. All of these implications have been tested extensively, often with surprising results. An overview of representative findings is presented below.

The first prediction of self-interference, regarding the necessity of multiple paths for individual quanta, was tested in a very creative way by Sillitto and Wykes [1]. They used an electric shutter to switch on and off the paths of a Young interferometer. Specifically, they made sure that the two paths were never open at the same time, yet both of them were switched several times before a photon had time to reach the detector. The rate of emission was low enough to have no more than one photon at a time in the apparatus ($10^6$ quanta/sec for a transit time of $10^{-8}$ sec), and interference was still observed. A possible vulnerability of this experiment is that different components of the same pulse could have accessed alternative paths, because the shutter was based on a two-channel polarizer. So, the actual distance between interfering units may have been smaller than implied by the rate of detection of photons. Nevertheless, the essential claim stands: no photon could have entered both paths simultaneously, and quanta from each path had to be temporally distinguishable. Furthermore, fringe visibility was shown to depend on phase-coherence, which was verified by adjusting the length of one of the paths. Consequently, in many cases where self-interference was thought to be the only viable explanation, some sort of remote interaction between non-overlapping photons must be at work. What could have been the real cause of interference in this case? The authors seem to believe that fast switching introduced an uncertainty in the time of detection, which restored indistinguishability. However, this does not explain the impact of phase coherence on fringe visibility (unless we are willing to assume that a particle, which definitely followed only one path, knew the exact length of the other, changing its trajectory without a physical cause).

A different way to test the requirement of two paths for every photon was done by Basano and Ottonello [2]. They used two lasers, well isolated from each other, ensuring that photons from each source could pass through only one of two slits. Moreover, the emission of photons from the two sources was not correlated in any obvious way.

The result was that interference did happen, even if none of the photons could access both paths. Again, this experiment does not rule out self-interference for all cases, but it does significantly restrain its area of likely validity. It should be mentioned that two-photon interference is very well investigated. In a sense, all instances of such interactions prove that photons can generate fringes from single distinguishable paths. However, the experiment of Basano and Ottonello is special because it used a double-slit setting, where self-interference is the standard interpretation. Moreover, the authors took special care to ensure the isolation of sources, in order to rule out a possible interaction between them. This test has a mutually reinforcing relationship with that of Silitto and Wykes, confirming the reality of interference among photons from separate paths, where self-interference alone was usually suspected.

An interesting addition to these results is the experiment of Santori *et al*. [3], which was done with two deterministic sources of single photons. They looked for bunching, not spatial fringes, but their results are surprising in many ways. For the purpose of this discussion, it is important to note that the time of emission of every photon was certain, and the path of every photon was certain when interference happened. This confirms the above comment that interference is possible between distinguishable photons, enough to create the appearance of self-interference when there is none. (To be fair, the authors did assume that photons became indistinguishable, but only because they saw interference). This experiment is also important because it violates the so-called energy-time complementarity. Fock states produce certainty about the number of photons, which should lead to uncertainty about phase. The fact that interference was observed suggests that uncertainty governs only our knowledge, not the actual interaction of individual photons. The experiment is also relevant for the next point of our discussion, by showing that interference between photons only happens within a limited interval between them. The shape of interference peaks indicated a gradual loss of fringe visibility with increasing distance between photons, similar to the likely effect of slowing down the rate of switching in the experiment of Silitto and Wykes.

As mentioned above, self-interference should not be sensitive to the size of the interval between consecutive photons. In fact, the evidence strongly suggests that fringe visibility does note correlate directly with intensity. Among the most popular references are the experiments of Parker, in which interference was produced at very low rates of detection ($10^5$ photons/sec) [4], [5]. Several recent set-ups have produced interference at much lower rates (down to one photon per second), but they are not published in major journals. Such results are almost taken for granted today. Nevertheless, some experiments also indicated the existence a threshold, beyond which interference vanished abruptly. This little known phenomenon was never explained in full. Dontsov and Baz [6] proposed that pulses become statistically independent at lower rates of emission, thereby losing coherence. They suggested that light-emitting atoms have weakly correlated states at lower energies. However, their own experiment showed that interference vanished regardless of the state of the source. A gray filter, which did not affect coherence behind the interference volume, destroyed interference when used to attenuate a stronger beam at the source. Moreover, light from independent sources does interfere, which shows that correlation at emission is not a necessary precondition for interference.

In a modern setting, with photons obtained via SPDC, Strekalov *et al*. [7] reported again that self-interference did not occur behind a double-slit. They were looking for "ghost interference" and needed very low intensities to accommodate coincidence windows greater than several picoseconds. These authors assumed that angular uncertainty in SPDC is too great for Young interference to occur. This was partly confirmed by a different experiment, in which narrow-band filters were used to restore fringe visibility [8]. Kim *et al.* discarded the idler beam (to remove "ghost" effects) and repeated the experiment at various intervals between consecutive photons. Their surprising discovery was that interference still vanished at the larger intervals between single photons. Specifically, no fringes were seen at intervals beyond one picosecond. This shows that in the experiment of Strekalov *et al*. [7], which used coincidence windows of several picoseconds, the fringes were not hidden. They had to be non-existent.

These recent experiments confirm the findings discussed by Dontsov and Baz, but leave open a different question. Why did they lose visibility in the neighborhood of one picosecond, if so many others saw fringes at average intervals greater than one microsecond? The likely answer is that the type of source does matter, but in a different way. Experiments like Parker's were done with attenuated laser pulses. They had a high probability of detecting single pulses, but could not control their width. If a pulse happens to be significantly wider than the arms of an interferometer, it can be detected alone and still overlap in space with a subsequent "wave packet". This possibility is less likely for SPDC sources, which produce very narrow single pulses. Consequently, the mentioned discrepancy is not a conceptual problem. Rather, it confirms the importance of the real interval between interacting photons. This point is further supported by another experiment, performed by Kim and Grice [9], in which photons arrived at a beam splitter well separated in time. Just like in other cases cited above, interference was detected, but only when pulses were relatively close to each other. Loss of fringe visibility with a gradual increase of the interval was well demonstrated, similar to the experiment of Santori *et al.* [3]. Moreover, phase coherence was crucial for the effect. Again, the details of the experiment make self-interaction implausible. Another unexpected element was that polarization had no effect on the outcome. Kim and Grice emphasized that the experiment was repeated many times with different polarization settings and their conclusion is firm. This brings us to the third prediction of self-interference, regarding path knowledge.

According to modern treatments, polarization acts as a path marker. This is in sharp contrast to classical interpretations, in which polarization is a transverse magnetic wave that acts on orthogonal planes to the direction of propagation. The classical model suggests that polarization should be relevant for the interference of simultaneous photons, which share the same plane. (This should be understood in approximate terms, because light photons seem to be more like pulses with real width. In other words, photons can also be "partially" simultaneous, when the intervals between pulses are smaller than their width). On the other hand, consecutive non-overlapping pulses must act on parallel planes, orthogonal to the direction of propagation. They can no longer interact magnetically. Their interference should be due to a different kind of waves (most likely, electric), which manifest an observable sensitivity to phase coherence. Kim and Grice are probably the first to show the irrelevance of polarization for temporally distinguishable photons. Yet, most of the experiments cited above confirmed the role of phase coherence for such settings. Consequently, the impact of knowledge on objective elements of interference must be overstated in mainstream interpretations. If polarization was merely active through its information content, it should have prevented interference in the quoted experiment, and phase-coherence should not have mattered. (If anything, the latter can only increase distinguishability).

Another element that could serve as path marker is wavelength. If a beam of blue photons were to be launched through one slit, and red photons through the other, which-path information would become instantly obvious. Therefore, interference between beams of different wavelength should be impossible. Yet, this is not the case. Such beams do not have stable fringes like coherent beams, and a certain degree of creativity is required for the detection of interference. Still, the effect was demonstrated with electromagnetic waves of different ranges, including with visible light. A good example is Louradour *et al.* [10]. This experiment resulted in clear fringes from two lasers with tunable average output wavelength near 568 nm. A picosecond streak camera was used to show even the fringe drift due to difference in wavelength. Granted, the detected photons were not attributed back to their paths after observation. However, they could only arrive along distinguishable paths. Moreover, some experiments have circumvented even this problem, as will be shown below.

The problem of distinguishability is very important, because it is often used to explain the fact that quanta display wave properties in some cases, but not others. This was also seen above. Nevertheless, classical waves also become locally indistinguishable when they overlap and interfere. They become distinguishable again after such interactions. So the fact that interference and paths are not observable simultaneously is not a non-classical phenomenon. Our limited knowledge is a consequence, not a cause of this process. In fact, if quanta are different from classical waves, it is because they can (!) be distinguished even during interference. This property was demonstrated in a large number of experiments that investigated ghost interference.

What is ghost interference? It is an interaction between non-overlapping entangled beams, which originate from adjacent sources. This interaction becomes detectable despite intervening modulators or guides, as long as the particles arrive at their separate detectors with coherent properties. The "ghost" in the definition probably comes from the fact that interference happens as if the two beams were superimposed, when they are clearly not. The size of the fringes can even be related to a virtual double-slit through their known proportionality to wavelength and distance to detector. In other words, this is an interference process between photons from separate paths, in which we have complete fringe visibility and complete path knowledge. For example, Ou *et al.* [11] have sent pairs of entangled photons through two independent unbalanced Michelson interferometers. Two fixed detectors were used to count the photons from each path. The long arms of the two interferometers were modulated simultaneously, such as to produce identical path differences in both of them. The experiment detected no interference in the singles count. However, observation in the coincidence count regime produced a cosine variation in the rates of detection, with a period equal to the wavelength of photons. It is remarkable that path difference (between the short and long arms) in each interferometer was too large for indistinguishability. Moreover, visibility was the highest when the coincidence window was smaller, such as to include only the cases when both entangled photons traveled along the short, or the long arms. Consequently, the crucial factor for the phenomenon was the simultaneous detection of photons that traveled along the modulated (long) paths. The most plausible interpretation is that entangled photons interacted with each other from remote non-intersecting paths.

These results are corroborated by the outcome of a similar experiment, performed (slightly earlier in the same year) by Kwiat *et al.* [12]. The only significant difference in this context was the fact that the two beams were guided in parallel through a single Michelson interferometer. Just like in the previous experiment, the authors detected a cosine variation. However, in their case the period of variation was twice shorter than the wavelength of the detected photons. The authors concluded that the wavelength of the pump pulse, which was used to produce pairs of entangled photons through downconversion, was the one that fit the results. This difference is troublesome for the assumption that path difference in the interferometer (between the short and long arms) was the crucial factor for interference. However, as mentioned above, this scenario was already ruled out by the other elements of the experiment. A more appealing explanation is to assume that photons from each pair interacted remotely, when they followed the same path. Therefore, the different period in the two experiments must be produced by the different angles of propagation of photons inside the interferometer. The modulation of path length prior to the final parallel alignment could have produced a spatial fringe-drift that can explain the cosine variation of the detection rates at fixed points.

A similar effect is observed when the paths are left unchanged, but the detectors are moved relative to each other. Again, the area of emission of the crystal (the source of entangled photons) is too large, and different source points must produce overlapping fringes. Therefore, interference should not be visible in the singles count. Instead, if one of the detectors is kept fixed, the other can be moved to scan all the detection area of the second beam. The fringes of coincident photons must have a high probability of sharing the same set of fringes, as if the two beams were part of a single Young interferometer. This mechanism implies that interference is necessarily due to the remote interaction of the simultaneous photons from their independent paths, somewhat like the interaction of temporally distinguishable photons with overlapping trajectories. In this case, the photons are simultaneous, but well separated in space. For confirmation, consider the experiment of Fonseca *et al.* [13] who did not divert the photons in any way, but merely placed a slit on the path of the idler beam and a wire on the path of the signal. They argued that the two elements produced a double slit when projected together (even though in reality they were separated). The authors obtained clear fringes on both paths (in separate iterations), but only in coincidence with fixed points on the other path. They also claimed that interference in their non-local double-slit could be described in terms of virtual path indistinguishability. Yet, this explanation is not necessary, considering that photons definitely followed independent non-intersecting paths, and the effect is known to happen even in the absence of such apertures (such as in [11, 12], discussed above). Consequently, photons must have interacted from perfectly distinguishable paths, just like in the previous two experiments. Moreover, this happened in a manner that allowed the unmistakable attribution of one path to each subset of fringes. In other words, path knowledge did not preclude the observation of interference. This undermines the third and final prediction of self-interference that was proposed for analysis in this presentation.

In light of the above, self-interference must be ruled out. None of its major implications were able to survive experimental assessment. Firstly, photons were shown beyond reasonable doubt to have well-defined single paths during double-slit interference. Secondly, they were proven to interact only within limited intervals from each other. Finally, interference revealed itself as a real phenomenon, independent from abstract informational processes. On the other hand, this conclusion is at odds with several major developments in modern quantum mechanics. In particular, it is incompatible with the hypothesis of quantum erasure, which is widely believed to explain many quantum processes, including self-interference. Consistency with quantum erasure has been reported in many experiments. Therefore, it must be given immediate attention, in order to avoid an interpretive inconsistency.

For the purpose of this discussion, double-slit quantum erasure is the most relevant, particularly in the demonstration of Walborn *et al.* [14]. The set-up is as follows: entangled photons, obtained in pairs through parametric downconversion, are detected independently. One photon (idler) is observed with a polarimeter, while the other (signal) is directed through a double-slit interferometer. Erasure is assumed when the two slits are covered by path markers, *i.e.* two symmetric circular polarizers with orthogonal fast axes. If the idler beam is detected without a polarizer, or with a diagonal plane polarizer, coincident signal photons produce no fringes. If the idler is detected with the polarizer parallel to the fast axis of one path marker, fringes are detected. If the idler is detected in parallel to the fast axis of the other signal path marker, antifringes are observed. The effect is real when the idler is detected before the signal, as well as afterwards. It looks as if a measurement of the idler erases the effect of path markers on the signal, even after the detection of the latter! Nevertheless, several complications spring out on closer inspection.

From a theoretical point of view, quantum erasure makes sense only if it really erases something. In this case, it has to erase the path information induced by the polarizers. Therefore, fringes must necessarily be due to Young interference (or, in this case, self-interference), and definitely not to other causes, such as direct signal-idler interaction (as seen in "ghost interference"). Yet, the latter explication is very likely, given the fact that the two beams are entangled. As a matter of fact, entanglement is crucial for this experiment, because the idler must acquire the polarization of the signal, before it can erase it at the moment of measurement. Though, having such a state gives it a high probability of being detected in parallel with the fast axis of the corresponding signal photon. In other words, coincidence measurements are just as likely to bring out the pairs of entangled photons that interacted directly. If so, then fringes would correspond to one path of the signal photons, and the anti-fringes to the other. Consequently, quantum erasure cannot be acceptable in this setting, unless signal photons are proven to produce Young interference at the same low rate of emission. (This could be done by repeating the experiment with one slit covered. Fringes should disappear for all settings, in contradiction to numerous other experiments with entangled beams). Moreover, it must also be proven that a known well-defined circular polarization has no effect on subsequent measurements of linear polarization for the same photon.

Quantum erasure also appears inconsistent, because it invokes the role of information, without demonstrating a significant correlation with its content. If such a correlation were real, fringes would reappear in the singles counts when path markers are manipulated to produce mixed settings with minimal path knowledge. However, a different experiment, performed by Schwindt *et al.* [15], suggested that they would not. The authors called this "non-erasing quantum erasure". Yet, their discovery adds credibility to the alternative interpretation, which claims that path markers are not crucial for interference in this setting. Moreover, as shown above, temporally distinguishable photons produce fringes regardless of their polarization [9]. Signal photons are distinguishable in this manner. So, if they were close enough, they might have produced fringes despite the path markers. The experiment proved the opposite: fringes did not appear for the total number of detection events, and were not likely to reappear in intermediate polarization settings either.

A separate problem is that quantum erasure is not yet completely defined. Originally, it was proposed as a fundamental phenomenon that could determine the properties of complementarity and self-interference [16]. If this were the case, interference fringes should be visible for the whole set of detected events in the case of early erasure (in contrast to delayed erasure, where the fringes can only be recovered in subsets). Such a difference between early and delayed erasure would leave no room for ambiguity about this phenomenon. However, Walborn *et al.* (just like many other experimenters) showed explicitly that fringes do not become observable for the entire set of detectable photons in such settings. One position of the idler polarimeter produced a small subset of fringes for the coincident signal photons. An orthogonal position produced a subset of antifringes. Together, the two sets can be added up to a Bell curve distribution, which shows no interference, just like in the case without polarized idler photons. Therefore, it is not necessarily true that the fringes were created by the act of observation. It is just as likely that they were simply hidden, only to be brought out with special detection techniques. If so, then erasure would hardly be more than an artifact of post-selection.

The report of Walborn *et al.* has another remarkable feature: fringes are obviously asymmetric in all the cases related to early detection of idler photons (even when there is no erasure involved), and much less so for the delayed detection. If erasure were an intervening phenomenon, there should have been a similarity among the cases without path markers, as well as among the cases with path markers. As it is, the time of detection of the idler appears to be the crucial element. This is much more consistent with the hypothesis of real photons, mentioned above with regard to distinguishability. From that point of view, delayed measurement of the idler allows it to interact with the signal until the latter is detected, thereby producing symmetric fringes. Early detection of the idler would leave the signal unescorted past the common volume, resulting in asymmetric fringes. A likely implication is that fringes should become altogether undetectable, if the idler were to be measured before signal photons reached the slits. Incidentally, this scenario is highly consistent with the experiment of Kim *et al.* that was discussed above [8]. As a reminder, the idler was discarded at emission in that experiment, and fringes vanished when signal photons were farther than one picosecond from each other. In the erasure experiment, coincidence windows of several picoseconds were used, which means that signal photons were too far from each other to interfere. The latter is also supported by the quoted experiment of Strekalov *et al.* [7], which used a very similar set-up and noticed no Young interference. Therefore, we must conclude that "double-slit quantum erasure" is actually direct interference between simultaneous signal and idler photons. The status of quantum erasure, in terms of experimental support, is not any stronger than that of self-interference, and therefore it can neither confirm nor predict it. The greater the number of experiments considered, the harder it is to defend non-realist models.

**Discussion.** The principle of complementarity holds that quanta may behave sometimes as particles, and sometimes as waves. In particular, interference is supposed to be amenable to analysis purely in terms of waves. Yet, the details presented above make such an interpretation impossible. In the same way in which particle collisions cannot explain the remote interaction of photons, waves cannot overlap if they are temporally distinguishable from each other, or if they can only access one of several available paths. Probability waves are not any more plausible, considering that path knowledge was shown to have no real impact on final outcomes. On the other hand, there are some elements that suggest the necessary presence of real waves. For example, remote interaction was shown to work only within limited spatial boundaries. Also, phase coherence among interfering photons was repeatedly found to be essential. Therefore, quantum interference must be interpreted in a way that involves both wave properties and particle properties simultaneously.

There are several models that can satisfy this requirement. They are known as pilot-wave models, as developed primarily by de Broigle and Bohm. However, earlier models were constrained by the need to accommodate the concept of self-interference. In light of the latest experimental record, this is not necessary. Rather, it is sufficient to assume that photons are pulses of real particles, which act as sources of real waves. This would make them similar to classical electrons that generate electric waves, or massive entities that produce gravitational waves. As shown elsewhere [17], this picture can provide a natural explanation for fringe build-up. Accordingly, individual photons can go through only one of several paths, and still produce fringes if they are within the range of each others' waves. They could also interact when they are temporally distinguishable, and even from separate non-intersecting paths.

The assumption that photons are sources of waves can explain all the features of interference that self-interaction could not. Moreover, it leads to an important prediction that is not made by other models. Hence, waves can only exhibit constructive or destructive interference inside the volume of interaction. Particles, on the other hand, must remain in clusters, if they are guided as such by the mutual interaction of their overlapping waves. Therefore, interference fringes for optical arrangements should be detectable in volumes that are beyond the regions of interaction, at least for a short period after passing through them. Remarkably, this phenomenon was observed by Hanbury Brown and Twiss (HBT) during their first demonstration of ghost interference [18]. In that classical set-up, photons traveled along a common path towards a half-silvered mirror. The beam was split 50-50, and two detectors were used to show correlations in the coincident count regime, as a function of their relative position. The interpretation proposed here is that concomitant photons, which propagated in parallel up to the beam splitter, interfered while being adjacent. After being redirected by the mirror, they left the interference volume, but maintained their momentum. This preserved the fringes, enabling the detection of points of maximum and minimum correlation during coincident measurements. The set-up of HBT is very similar to recent experiments with entangled beams, the difference being that photons were redirected after their interaction. It is important to note that fringes persisted even when detectors were not equidistant from the mirror. This confirms the hypothesis of photon clustering beyond the interference volume, and also provides a classical explanation for the appearance of delayed quantum erasure.

Self-interaction was the main obstacle for realist interpretations of the double-slit experiment. As shown by Khrennikov in his work on p-adic models of probability [19], classical processes could reproduce quantum distributions if detection events were not independent from each other. Yet, non-classical mechanisms still seemed to be required if self-interference was real. The alternative proposed above is free from such constraints, but also limits the validity of the formalism to well-defined physical contexts. Photons from different paths must be sufficiently close to each other, if they are to produce fringes. This is already confirmed by experiment, as shown, but it is not a generally accepted phenomenon. Additional testing is necessary, in order to remove all doubts. For example, entangled non-overlapping coherent beams should produce fringes in the singles count regime (contrary to current expectations), if narrow slits were placed as close as possible to their sources. Also, interference should be detected with a double-slit interferometer on the path of one beam, when the other is discarded, provided the rate of emission is increased beyond the thresholds mentioned above. Furthermore, entangled beams should not produce visible fringes even in the coincidence count regime, if they were redirected at 180 degrees to each other immediately after emission, and detected at larger distances from their sources. This prediction extends only to the so-called position correlations. It would not apply to spin correlations, which survive separation, because they are not produced by direct interaction among particles after emission. (See [20] for a discussion of the EPR paradox, from this realist perspective).

The hypothesis of particulate wave-producing photons leads to interesting connections with other concepts in modern physics. It is compatible with a brane-world scenario, in which localized particles engage in mediated interaction [21]. It explains the frequency of de Broigle matter waves in terms of corresponding numbers of elementary wave-sources, detectable as macro-particles. It also provides a classical mechanism for the operation of Maxwell's equations, by showing that electric and magnetic components are produced simultaneously in discrete steps by photons. Specifically, electric fields must be produced as longitudinal 3D waves, in order to explain the inverse square law and the role of phase coherence during interference; magnetic fields would be simultaneously produced as 2D transverse waves, which predict the features of polarization. The manipulation of fields must be due to the effect of some particles on other particles. Magnetic and electric fields are not expected to generate each other in vacuum, in the absence of matter. Furthermore, this approach is in good agreement with Huygens' model of wave propagation. It also provides a natural explanation for the directionality and quantized nature of electromagnetic waves.

On the other hand, the implications of special relativity are not supported. Our model explains the constant speed of light in terms of general relativity, via the effect of gravity on the motion of photons (similar to the interpretation of Maxwell's fields by Theocharis [22]). Such a position is quite plausible, considering that numerous tests of invariance contain persistent artifacts, which are not anticipated by special relativity (see, for example, [23]). These effects can probably be explained naturally by reference to the Sagnac effect, produced by the motion of terrestrial labs around Earth's center of gravity, and other phenomena that fit general relativity. Additional arguments are provided by the appearance of anomalous acceleration of deep space probes (as reported, e.g., in [24]), which might be removed by switching to a heliocentric frame of reference. This enhances the credibility of a likely preferred status for general relativity in Nature, insofar as the gravity of the Sun is predominant in the interplanetary medium.

Perhaps the most surprising element of this interpretation is the fact that the speed of light is the speed of particles of light, not of electric and magnetic waves. The propagation of the latter explains the emergence of scalar potentials, but it cannot be instantaneous. These are, after all, real waves spreading inside a real medium (the brane). They must have a finite speed, which may or may not be the speed of light. In fact, they cannot always be expected to have the same speed, because they must violate Lorenz invariance (unlike their sources). A brane is expected to be a relatively stable cosmic medium, with a preferred frame of reference for the speed of its waves. This conclusion also implies that the interpretation of the Michelson-Morley experiment should be reconsidered. If photons are assumed to be particles, their speed has no effect on our conclusions about the speed of associated waves. The latter must be verified by testing the speed of propagation of electrostatic or magnetostatic pulses in opposite directions.

**Acknowledgments**. Serafino Cerulli-Irelli recommended many published reports that were essential for this analysis. Frank Zimmermann, Sheldon Goldstein, Marlan Scully, Carlos Monken, Steve Walborn, Yoon-Ho Kim, Paul Kwiat, Andrei Khrennikov, Greg Jaeger, Jack Sarfatti, Andrei Akhmeteli and Shahriar Afshar provided stimulating feed-back and helpful criticism.